\def\year{2020}\relax
\documentclass[letterpaper]{article} 
\usepackage{aaai_authorkit20/aaai20}  
\usepackage{times}  
\usepackage{helvet}  
\usepackage{courier}  
\usepackage[hyphens]{url}  
\usepackage[english]{babel}
\usepackage[utf8]{inputenc}
\usepackage{algorithm}
\usepackage[noend]{algpseudocode}
\usepackage{graphicx}  
\usepackage{amsmath}
\usepackage{booktabs}
\usepackage[margin=1in]{geometry} 
\usepackage{amsmath,amsthm,amssymb,amsfonts, fancyhdr, color, comment, graphicx, environ}
\usepackage{xcolor}
\usepackage{mdframed}
\usepackage[shortlabels]{enumitem}
\usepackage{indentfirst}
\usepackage{makecell}
\usepackage{balance}

\renewcommand{\qed}{\quad\qedsymbol}

\binoppenalty=\maxdimen
\relpenalty=\maxdimen
\usepackage{xcolor}

\usepackage{amssymb}
\title{Optimizing Vulnerability-Driven Honey Traffic Using Game Theory}
\author{Iffat Anjum\textsuperscript{\rm1}, Mohammad Sujan Miah\textsuperscript{\rm2}, Mu Zhu\textsuperscript{\rm1}, Nazia Sharmin\textsuperscript{\rm2}, \\ \Large \textbf{Christopher Kiekintveld\textsuperscript{\rm2},  William Enck\textsuperscript{\rm1}, Munindar P Singh\textsuperscript{\rm1}}\\ 
\textsuperscript{\rm1} North Carolina State University,\\ Raleigh, NC 27695\\
\{mzhu5, ianjum, mpsingh, whenck\}@ncsu.edu\\
\textsuperscript{\rm2} The University of Texas at El Paso, \\El Paso, TX 79968\\
\{msmiah@miners., nsharmin@miners., cdkiekintveld@\}utep.edu}

\date{}
\begin{document}

\maketitle
\pagestyle{plain}
\thispagestyle{plain}

\begin{abstract}
Enterprises are increasingly concerned about adversaries that slowly and  deliberately exploit resources over the course of months or even years.
A key step in this kill chain is network reconnaissance, which has historically been active (e.g., network scans) and therefore detectable.
However, new networking technology increases the possibility of \emph{passive} network reconnaissance, which will be largely undetectable by defenders.
In this paper, we propose \emph{Snaz}, a technique that uses deceptively crafted honey traffic to confound the knowledge gained through passive network reconnaissance.
We present a two-player non-zero-sum \emph{Stackelberg} game model that characterizes how a defender should deploy honey traffic in the presence of an adversary who is aware of \emph{Snaz}.
In doing so, we demonstrate the existence of optimal defender strategies that will either dissuade an adversary from acting on the existence of real vulnerabilities observed within network traffic, or reveal the adversary's presence when it attempts to unknowingly attack an intrusion detection node.

\end{abstract}

\section{Introduction} \label{sec:intro}

Advanced Persistent Threats (APTs) are a significant concern for enterprises.
In such scenarios, advanced adversaries take slow and deliberate steps over months and even years to compromise critical resources (e.g., workstations and servers) in a network.
A key step in the kill chain of APTs is reconnaissance.
Historically, reconnaissance is largely active, for example using network port scanning to identify which hosts are running which services.
In response, many enterprises closely monitor their networks for scanning attacks.

Simultaneously, Software Defined Networking (SDN) technology is emerging as a powerful primitive for enterprise network security.
SDN offers a global perspective on network communications between hosts.
It can be used as an enhanced tool to identify network scanning, provide flexible access control to mitigate attackers bypassing defenses such as firewalls, and even prevent spoofing.
However, the increased functionality within network elements (e.g., switches) makes them a target for attack.
A compromised SDN switch is particularly dangerous, because it can perform reconnaissance passively.
As a result, defenders may have little to no signal that an APT is in process.

We propose \emph{Snaz} ("snag and zap") to address the threat of passive network reconnaissance.
\emph{Snaz} uses honey traffic: fake flows deceptively crafted to make a passive attacker think specific resources (e.g., workstations and servers) exist and have specific un-patched, vulnerable software.
\emph{Snaz} assumes the adversary knows about the possibility of honey traffic and uses game theory to characterize how best to send honey traffic.
For doing so, we demonstrate how a defender can either successfully deflect or deplete an adversary using optimal amount of honey traffic.

\emph{Snaz} models this defender-attacker interaction as a two-player non-zero-sum Stackelberg game.
In this game, the defender sends honey traffic to confound the adversary's knowledge.
However, if the defender sends too much honey traffic, the network may become overloaded.
In contrast, the adversary wishes to act on information obtained using passive reconnaissance (e.g., a banner string indicating a server is running a vulnerable version of Apache).
However, if the adversary acts on information in the honey traffic, it will unknowingly attack an intrusion detection node and be discovered.
Thus, the game presents an opportunity to design an optimal strategy for defense. We make the following primary contributions:


\begin{table*}[t]
    \centering
    \caption{Example of types of information used for attacks}
    \label{tab:attacks}
    \begin{tabular}{p{4.5cm}|p{5.5cm}|p{4.5cm}}
    \Xhline{2\arrayrulewidth}
       \textbf{Target Type}  & \textbf{Analysis Space} & \textbf{Examples}\\ \hline 
       {Fingerprinting OS} & {TTL, Packet Size, DF Flag, SackOk, NOP Flag, Time Stamp} & {Windows 2003 and XP} \\ \hline
       {Server software, version, \newline service type} & {Default banners} & {Apache HTTP 2.2, \newline Windows Server 2003} \\ \hline
       {Network topology, \newline forwarding logic} & {Flow-rule update frequency, controller-switch communication} & {Lack of TLS adoption, modified flow rules} \\ \hline
       {Employee Credentials, \newline personal information} & {Server-client traffic header and data} & {HTTP traffic, HTTPS traffic \newline with weak TLS/SSL} \\ 
    \Xhline{2\arrayrulewidth}
    \end{tabular}
\end{table*}

\begin{itemize}
    \item \textit{We propose \emph{Snaz}, a technique of using honey traffic to mitigate the threat of passive network reconnaissance.}
    This honey traffic has the potential to dissuade an adversary from acting on evidence of real vulnerabilities or more quickly reveal its existence within the network.
    \item \textit{We model \emph{Snaz}'s deceptive defense using a two-player non-zero-sum Stackelberg game.} We present an algorithm for finding the optimal strategy for deploying honey flows that is fast enough to be used for realistic networks.
    \item \textit{We present an empirical evaluation} of the performance of our game model solutions under different conditions, as well as the scalability of the algorithm and some useful properties of the optimal solutions.
    \item \textit{We emulate \emph{Snaz} in Mininet\cite{de2014using}, also showing the network overhead that results from honey traffic.}
\end{itemize}

\section{Background and Related Work}
\label{sec:Background}



Enterprise network administrators are increasingly concerned with Advanced Persistent Threats (APTs) where adversaries first obtain a small foothold within the network and then stealthily expand their penetration over the course of months, sometimes years.
The past decade has provided numerous examples of such targeted attacks, e.g., Carbanak~\cite{Carbanak}, OperationAurora~\cite{OperationAurora}.
Such attacks require significant planning.
Initially, adversaries identify attack vectors including 1)~vulnerable servers or hosts, 2)~poorly configured security protocols, 3)~unprotected credentials, and 4) vulnerable network configurations.
To do so, they leverage network protocol banner grabbing, active port scanning, and passive monitoring~\cite{kondo2014penetration,passiveactive:2007}.
Examples of different types of desired information and corresponding attacks are shown in Table~\ref{tab:attacks}.

Software Defined Networking (SDN) has the potential to address operational and security challenges large enterprise networks~\cite{Levin:2014:Panopticon}.
They provide flexibility to programmatically and dynamically re-configuring traffic forwarding within a network~\cite{McKeown:2008:OpenFlow} 
and provide opportunities for granular policy enforcement~\cite{Kim:2013:SDN}.
However, these more functional network switches form a large target for attackers as they can provide a foothold to perform data plane attacks using advanced reconnaissance and data manipulation and redirection~\cite{Markku:2014:Spook}.
Using one or more compromised switches, an adversary can learn critical information to mount attacks, including network topology and software and hardware vulnerabilities~\cite{Benton:2013:OVA,jero2017researchBEADs}.  

Deception is an important tactic against adversary reconnaissance, and there have been a variety of different approaches that apply game-theoretic analysis to cyber deception~\cite{pawlick2019game}. Many of these previous works have focused on how to effectively use honeypots (fake systems) as part of a network defense~\cite{carroll2011game,wagener2009self,pibil2012game,kiekintveld2015game}. This has included work on signaling games where the goal is to make real and fake systems hard to distinguish~\cite{miah2020concealing}.  Work on security games (including games modeling both physical and cybersecurity) focuses on deception to manipulate the beliefs of an attacker~\cite{yin2013optimal,an2011refinement,horak2017manipulating,thakoor2019general}.
Another research \cite{schlenker2018deceiving} proposes a game model of deception in the reconnaissance phase of an attack, though they do not consider honey flows. Stackelberg game models have been used to find optimal strategies for cyber-physical systems~\cite{feng2017stackelberg}.


\section{Overview}
\label{sec:Assumptions}

This paper proposes \emph{Snaz}, a deception system designed to mislead or delay the passive reconnaissance by an adversary.
Snaz provides this deception using \emph{honey traffic} that is precisely controlled by the defender.
We now give a high level overview of the system and threat model.

\subsection{Snaz}

Snaz uses honey traffic to mislead adversaries using passive network reconnaissance.
This deception consists of network flows with fake information, which we call \emph{honey flows}.
Traditionally, a network flow is defined as a 5-tuple: source IP, source port, destination IP, destination port, and protocol (e.g., TCP).
For simplicity, we assume honey flows include network flows in both directions to simulate real network communication.

Honey flows can fake information in network flow identifiers.
For example, a honey flow can attempt to make the adversary believe a non-existent host has a specific IP address, or a host is running a server on a specific port.
Due to the flexible packet forwarding capabilities of SDN, the defender can route honey flows through any path it chooses, e.g., to tempt an adversary that has compromised a switch on a non-standard path.
Honey flows can fake information in the packet payload itself.
For example, network servers often respond with a banner string indicating the version of the software, and sometimes even the OS version of the host.
Attackers often use this banner information to identify unpatched vulnerabilities on the network.
Honey flows can simulate servers with known vulnerabilities, making it appear as if there are easy targets.
If at any point the adversary acts on this information (i.e., connects to a fake IP address), Snaz will redirect the traffic to an intrusion detection node.
Since the intrusion detection node does not normally receive network connections, the existence of any traffic directed towards it indicates the presence of an adversary on the network.

\begin{figure}[t]
  \centering
 \includegraphics[width=\linewidth]{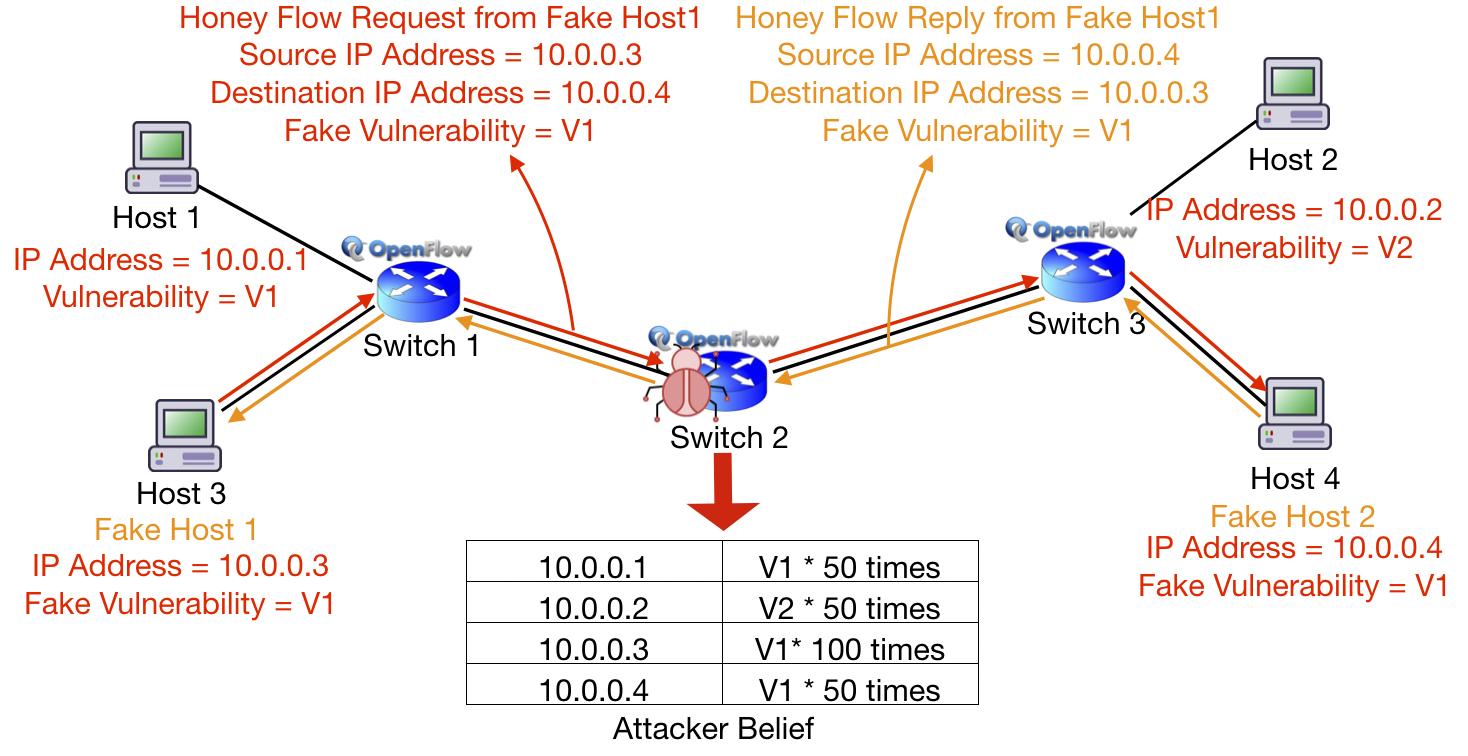}
  \caption{Snaz uses honey traffic to mislead adversaries.}
  \label{example1}
\end{figure}

Figure~\ref{example1} shows a simplified example of honey flows causing an adversary to update its belief.
The figure shows two real hosts: $Host~1$ with vulnerability type $V1$, and $Host~2$ is with vulnerability type $V2$. 
The adversary has compromised $Switch~2$ and observes all packets passing through it.
Without honey traffic, the adversary can easily identify the vulnerabilities on the hosts (e.g., via banner strings) and attack them.
In the figure, Snaz simulates the existence of two fake hosts ($Host~3$ and $Host~4$) using honey traffic.
If the adversary is unaware of Snaz, it will probabilistically attack either $Host~3$ or $Host~4$ and be quickly detected.
However, if the adversary \emph{is} aware of Snaz (the scenario we consider in this paper), it must keep track of what it believes is real verses fake information.
How the defender and attacker act is the crux of our game-theoretic model in Section~\ref{sec:Game}.

\subsection{Threat Model and Assumptions}

The adversary's goal is to compromise networked resources, e.g., workstations and servers, without detection.
The adversary does not know what hosts are on the network or which hosts have vulnerabilities.
It must discover vulnerable hosts using network reconnaissance.
The adversary assumes the defender has deployed state-of-the-art intrusion detection systems that can identify active network reconnaissance such as network port scanning.
However, we assume the adversary has gained a foothold on one or more network switches (an upper-bound of which is defined by the model).
Using this vantage point, the adversary is able to inspect all packets that flow through the compromised switches.
In doing so, it can learn (1) network topology and which ports servers are listening to by observing network flow identifiers, (2) about the installed software versions by observing server and client banner strings.
We assume the adversary can map between banner strings and known vulnerabilities and their corresponding exploits.
We conservatively assume this mapping can occur on the switch, or can be done without the knowledge of the defender.
The adversary also has the capability to initiate new network flows from the switch while forging the source IP address, as response traffic will flow back through the compromised switch and terminate as if it was delivered to the real host.
Finally, we assume the adversary is rational and is aware of the existence of Snaz and that honey traffic may be sent to fake hosts.
However, the adversary does not know the specific configuration of Snaz, such as the distribution of honey traffic.

We assume the defender's network contains real hosts with exploitable vulnerabilities.
The defender is aware of some, but not all of these vulnerabilities.
For example, the defender's inventory system may indicate the existence of an unpatched and vulnerable server, but due to production requirements, the server is not yet patched.
We further assume the defender can identify valuations of each network asset and approximate the valuation of the assets to the attacker (e.g., domain controllers that authenticate users are valuable targets).
We assume that the SDN controller and the applications running on the controller are part of the trusted computing base (TCB).
We further assume that the communication between the SDN controller and uncompromised switches is protected and not observable to the adversary (e.g., via SSL or an out-of-band control network).
As a result, the adversary cannot alter the controller configuration or forwarding logic of uncompromised switches.
Finally, we assume that the network utilization is not near maximum capacity during normal operation.
However, exceeding honey traffic may cause congestion and cause network degradation.

\section{Game Model}
\label{sec:Game}

An important question we must answer to deploy \emph{Snaz} effectively is how to optimize the honey traffic created by the system, including how much traffic to create of different types.  This decision must balance many factors, including the severity of different types of vulnerabilities, their prevalence on the network, and the costs of generating different types of honey flows (e.g., the added network congestion). 
In addition, a sophisticated APT attacker may be aware of the possible use of this deception technique, so the decisions should robust against optimal responses to honey traffic by such attackers. Finally, we note that many aspects of the environment can change frequently; for example, new zero day vulnerabilities may be discovered that require an immediate response, or the characteristics of the real network traffic may change. Therefore, we require a method for making fast autonomous decisions that can be adjusted quickly. 

We propose a game theoretic model to optimize the honey flow strategy for \emph{Snaz}. Our model captures several of the important factors that determine how flows should be deployed against a sophisticated adversary, but it remains simple enough that we can solve it for realistic problems in seconds (see Section~\ref{sec:Evaluation} for details) allowing us to rapidly adapt to changing conditions.  Specifically, we model the interaction as a two-player non-zero-sum Stackelberg game between the defender (leader) and an attacker (follower) where the defender (\emph{Snaz}) plays a mixed strategy and the attacker plays pure strategy. This builds on a large body of previous work that uses Stackelberg models for security~\cite{tambe2011security}, including cyber deception using honeypots~\cite{pibil2012game}. 

\begin{table}[htb]
\centering
\caption{Game Notation}
\label{tab:notation}
\begin{tabular}{|l p{2.3in}|}\hline
$d$ & defender\\
$a$ & attacker\\
$V_i \in V$ & set of $i$ types of vulnerabilities in the network\\
$R_i$ & number of real flows indicating $V_i$\\
$H_i$ & upper bound on the number of honey flows indicating $V_i$\\
$\Phi_{ij}$ & action of selecting $j\in[0,H_i]$ honey flows for $V_i$\\
$\Phi$ & defender's mixed strategy as the marginal probabilities over $\{\Phi_{i0},\dots,\Phi_{iH_i}\}$\\
$C_i$ & cost of creating each honey flow that indicates $V_i$\\

$\upsilon_i^{a,r} \; \upsilon_i^{a,h}$  & the value the attacker gains from attacking a real or fake flow of type $V_i$\\
$\upsilon_i^{d,r} \; \upsilon_i^{d,h}$  & the value the defender loses from an attack against a real or fake flow of type $V_i$\\

$a_i$ & denotes the action of attacking a flow of type $V_i$ where $a_0$ is the no-attack action, yielding 0 payoff \\
     &\\
     \hline
     \end{tabular}
\end{table}

We now formally define the strategies and utilities of the players using the notation listed in Table~\ref{tab:notation}. We assume that the defender is using \emph{Snaz} as a mitigation for a specific set of $i$ vulnerabilities that we label $V_i$. Every flow on the network indicates the presence of at most one of these types of vulnerabilities in a specific host. The real network traffic is characterized by the number of real flows $R_i$ that indicate vulnerability type $V_i$. The pure strategies for \emph{Snaz} are vectors that represent the number of honey flows that are created that indicate each type of vulnerability $V_i$; we write $\Phi_{ij}$ to represent the marginal pure action of creating $j$ flows of type $V_i$. These fake flows do not need to interact with real hosts; they can advertise the existence of fake network assets (i.e., honeypots). The defender can play a mixed strategy that randomizes the number of flows of each type that are created, which we denote by $\Phi$. To keep the game finite we define the maximum number of flows that can be created of each type as $R_i$. The attacker's pure strategy $a_i$ represents choosing to attack a flow of type $V_i$, or not to attack. We assume the attacker cannot reliably distinguish real flow and honey flow, so an attack on a specific type corresponds to drawing a random flow from the set of all real and fake flows of this type.

The utilities for the players depend on which vulnerability type the attacker chooses, as well as on how many real and honey flows of that type are on the network. An attack on a real flow will result in a higher value for the attacker than on a honey flow of the same type, and vice versa for the defender. Specifically, if the attacker chooses type $V_i$, it gains a utility $\upsilon_i^{a,r}$, which is greater than or equal to the value for attacking a honey flow of the same type $\upsilon_i^{a,h}$ (which may be negative or 0). We assume that this component of the utility function is zero-sum, so the defender's values are $\upsilon_i^{d,r} = - \upsilon_i^{a,r} $ and $\upsilon_i^{d,h} = - \upsilon_i^{a,h}$. The defender's utility function includes a cost term $C_i$ that models the marginal cost of adding each additional flow of type $V_i$ (for example, the additional network congestion which can vary depending on the type of flow). If the defender plays strategy $\Phi$ and the attacker attacks the $V_i$, the defender's expected utility is defined as follows:

\begin{equation}\label{eq:defender-EV}
U^d(\Phi, i) = P_i^r \upsilon_i^{d,r} + (1-P_i^r) \upsilon_i^{d,h} - C^h\\
\end{equation}

Here, $P_i^r$ denotes the probability of attacking a vulnerability of type $V_i$ which can be calculated as follows:
\begin{equation*}
RSMN (mc, P) = \sum_{j \in \{0,\dots,H_i\}}\Phi_{ij}(R_i/(j+R_i))\\      
\end{equation*}

The overall cost $C^h$ for playing $\Phi$ is given by Equation~\ref{eq:defender-EV}:
\begin{equation*}\label{eq:defender-cost}
C^h =  \sum_{i \in V} \sum_{j \in \{0,\dots,H_i\}}(\Phi_{ij} \times j \times C_i)\\
\end{equation*}

Analogously, for the attacker the expected utility is given by:
\begin{equation}\label{eq:attacker-EV}
U^a(\Phi, i) = P_i^r \upsilon_i^{a,r} + (1-P_i^r) \upsilon_i^{a,h}
\end{equation}

 
\subsection{\emph{Snaz} Game Example}
Consider a network with two types of vulnerabilities.
Let the values be $\rm \upsilon^{a,r} = (10, 20)$, and $\rm \upsilon^{a,h} = (-5, -10)$, and the cost of creating honey flow indicating each type of vulnerability is $\rm C = (1, 0.5)$.
The total number real flows indicating each type of real vulnerability is $\rm R = (5, 5)$, and the upper bound on honey flows is $\rm H = (2, 3)$.
Thus at most two honey flows of $1^{st}$ vulnerability type and three honey flows of $2^{nd}$ vulnerability type can be created. 
Now, consider if the defender plays the following strategy $\Phi$:

\begin{figure}[h]
\includegraphics[width=0.7\columnwidth]{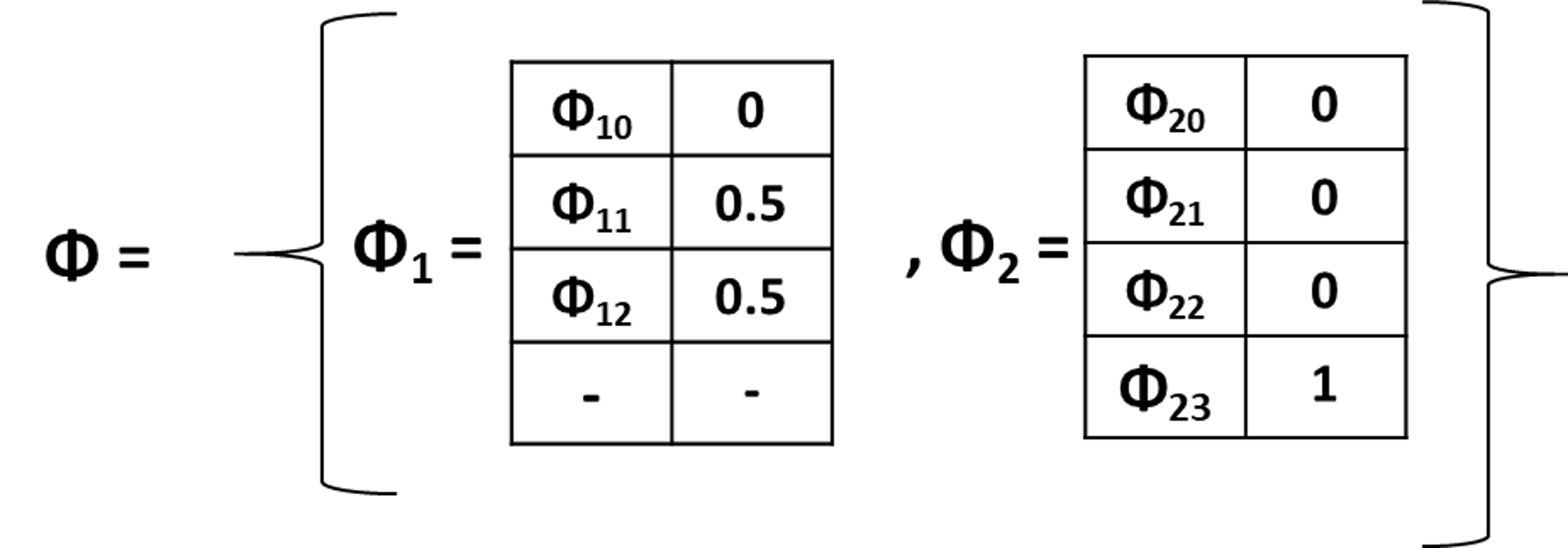}
\centering
\end{figure}

In $\Phi$ the defender creates one honey flow $50\%$ of the time and two honey flows $50\%$ of the time with type 1 vulnerability.
The defender also creates three honey flows $100\%$ of the time type 2 vulnerability. 
The attacker's best response is to attack vulnerability type 2 with expected utility $\rm U^a(\Phi, 2) = 8.75$, and the defender utility is $\rm U^d(\Phi, 2) = - 11.75$. 

\subsection{Optimal Defender's Linear Program}
Our objective is to compute a Stackelberg equilibrium that maximizes the defender’s expected utility, assuming that the attacker will also play the best response. 
To determine the equilibrium of the game,  we formulate a linear program (LP) where the attacker's pure strategy $a$ is a binary variable.  
We create a variable for each defender's pure strategy $\Phi_{i,j}$, the action of creating  $\rm j$ honey flows for $V_i$.  
The following LP computes the defender's optimal mixed strategy for each type of vulnerability under the constraint that the attacker plays a pure-strategy best response:

\begin{equation}\label{eq:defmax}
\max_{i \in V} \;\; U^{d}(\rm \Phi,  i) \;a_i
\end{equation}
\begin{equation*}\label{eq:range}\begin{split}
s.t.\;\;\;
a_i \in \{0,1\}, \;\; \Phi_{ij} \in [0,1]
\end{split}
\end{equation*}
\begin{equation} \label{eq:suchthat}
 U^{a}(\rm \Phi, i) \;a_i \geq
 U^{a}(\Phi, i')\; a_{i} 
\quad \forall\;i,i' \in V \;
\end{equation}
\begin{equation}\label{eq:def-strategy}
\sum_{ j\in\{0,\dots,H_i\}} \Phi_{ij} = 1 \quad \forall \Phi_i \in \Phi \; 
\end{equation}
\begin{equation}\label{eq:at_total}
\sum_{i \in V} a_i = 1  
\end{equation}

In the above formulation the unknown variables are the defender's strategy $\{\rm \Phi_{i0}, \dots, \Phi_{iH_i}\}$ for each $\Phi_i \in \Phi$ and the attacker's action $a_i$. 
Equation~\ref{eq:defmax} is the objective function of the LP that maximizes the defender’s expected utility.
The inequality in Equation~\ref{eq:suchthat} ensures that the attacker plays a best response.
Finally, Equation~\ref{eq:def-strategy} forces the defender strategy to be a valid probability distribution.
\section{Simulations and Model Analysis}
\label{sec:Evaluation}
We now present some results of simulations based on our game-theoretic model, as well as with an initial implementation of honey flow generation in an emulated network environment. We show that the game theory model can produce solutions that improve over simple baselines, and can be calculated fast enough to provide solutions for realistic networks.  We also examine how the optimal solutions change based on the parameters of the model to better understand the structure of the solutions and the sensitivity to key parameters of the decision problem. 

\subsection{Preliminary Testbed Evaluation}
\label{sec:Mininet}
\begin{figure}[htb!]
\includegraphics[width=1.0\columnwidth,height=0.5\columnwidth]{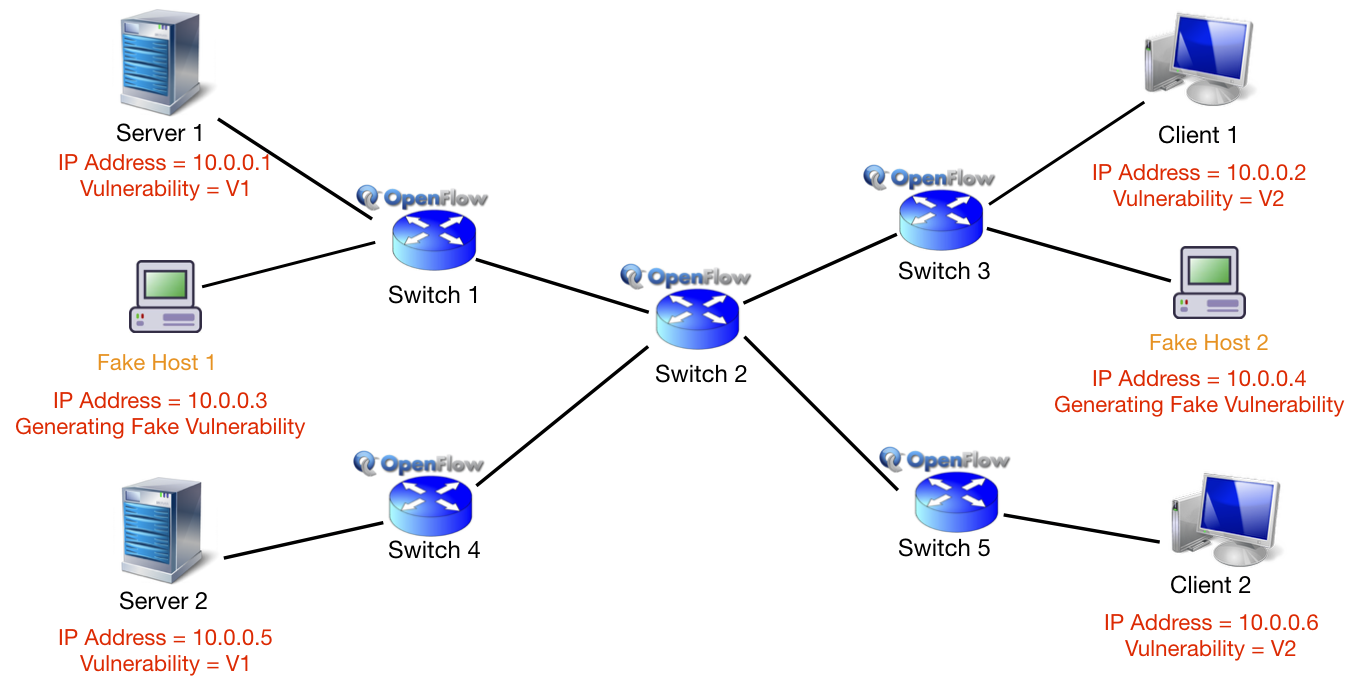}
\centering
\caption{Simple Network Topology used for Mininet Simulation}
\label{fig:mininet_topo}
\end{figure}

We have constructed a preliminary honey flow system in the emulated environment of \emph{Mininet}~\cite{Mininet}. 
We want to show the possibility of generating plausible \emph{honey traffic} in an emulated network. 
And also want to evaluate the effects of the deception in a more realistic context. 
We work on a small topology shown in Figure~\ref{fig:mininet_topo}. 
As shown in the figure, we consider four real and two fake hosts.
Some additional parameters:
\begin{itemize}
    \item \emph{Client 1} connects with \emph{Server 1}, while \emph{Client 2} connects with \emph{Server 2}. In the simulation, each of these two clients send 500 packets to the servers and receive corresponding replies.
    \item The \emph{fake clients} are connected with the system and can send packets with fake vulnerabilities to each other.
    \item All the links in our simulation are 1 MBit/s bandwidth, 10ms delay and $2\%$ probability loss.
    \item The values assigned to the servers are $2$ and real clients have a value of $1$ for both of attacker and defender.
\end{itemize}

\begin{figure}[htb!]
\centering
\includegraphics[width=1.1\columnwidth,height=0.5\columnwidth]{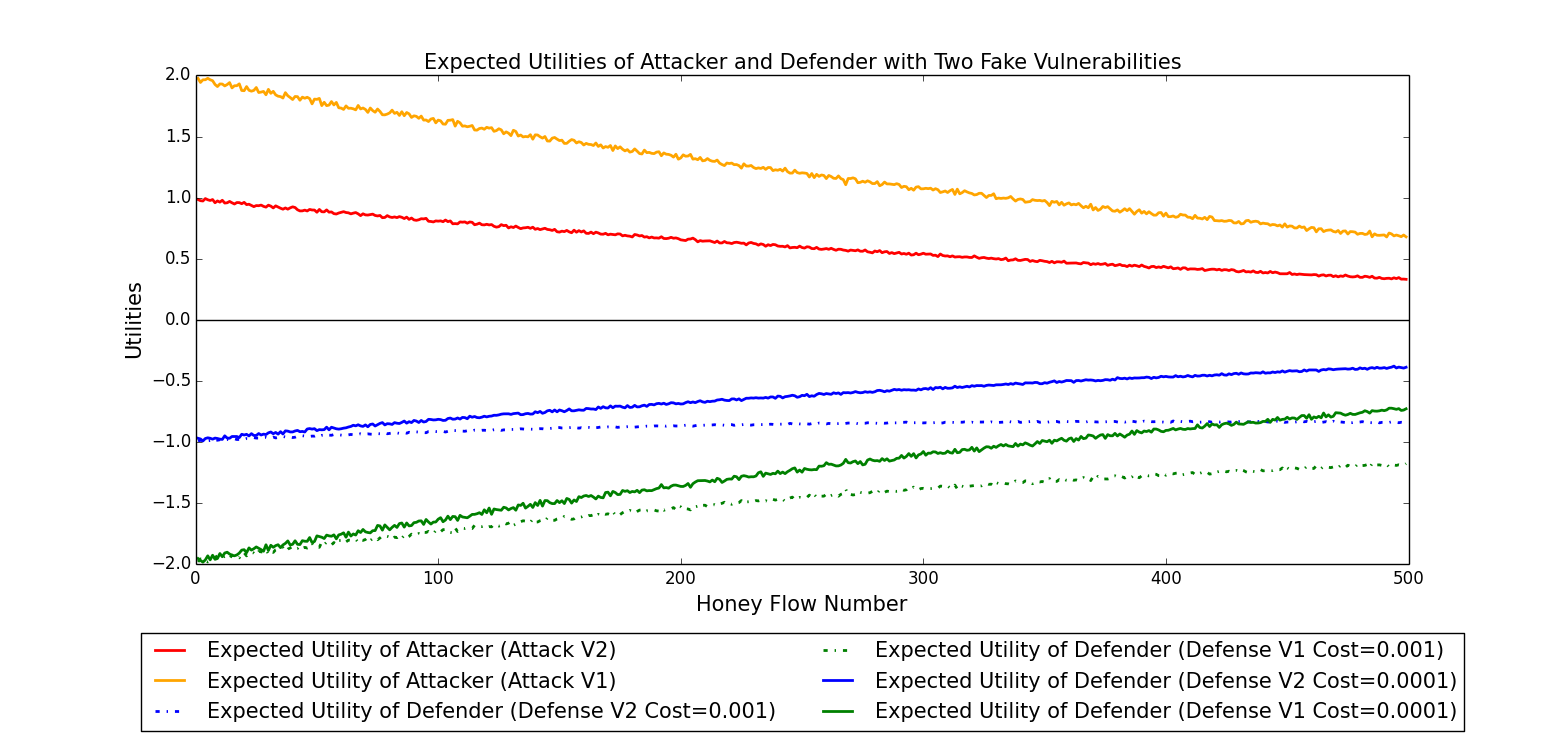}
\caption{Defender and attacker utility.}
\label{fig:cost_vul_2_2}
\end{figure}

We visualize the simulation results with two types of vulnerabilities in Figure~\ref{fig:cost_vul_2_2}. In this test, we increase the amount of honey flows from the fake clients from 0 to 500 packets. The honey flows from fake client 1 are with vulnerabilities type 1, while fake client 2 sends honey flows with vulnerabilities type 2. Besides, we experiment with different costs of honey flow generation. The dashed lines in Figure~\ref{fig:cost_vul_2_2} shows the expected utilities when generate 1 honey flow with cost 0.001 and solid lines represent that defender has to endure cost 0.0001 to generate each honey flow.

This simulation does not use the game theoretic optimization. 
However, we can still get the idea that increasing the number of honey flows remarkably reduces the effectiveness of adversarial efforts.
Meanwhile, we can also see that the cost of honey flow generation significantly affects the defender utility.


\subsection{\emph{Snaz} Game Theory Solution Quality} 

\begin{figure}[ht!]
\includegraphics[width=1.0\columnwidth,height=1.0\columnwidth]{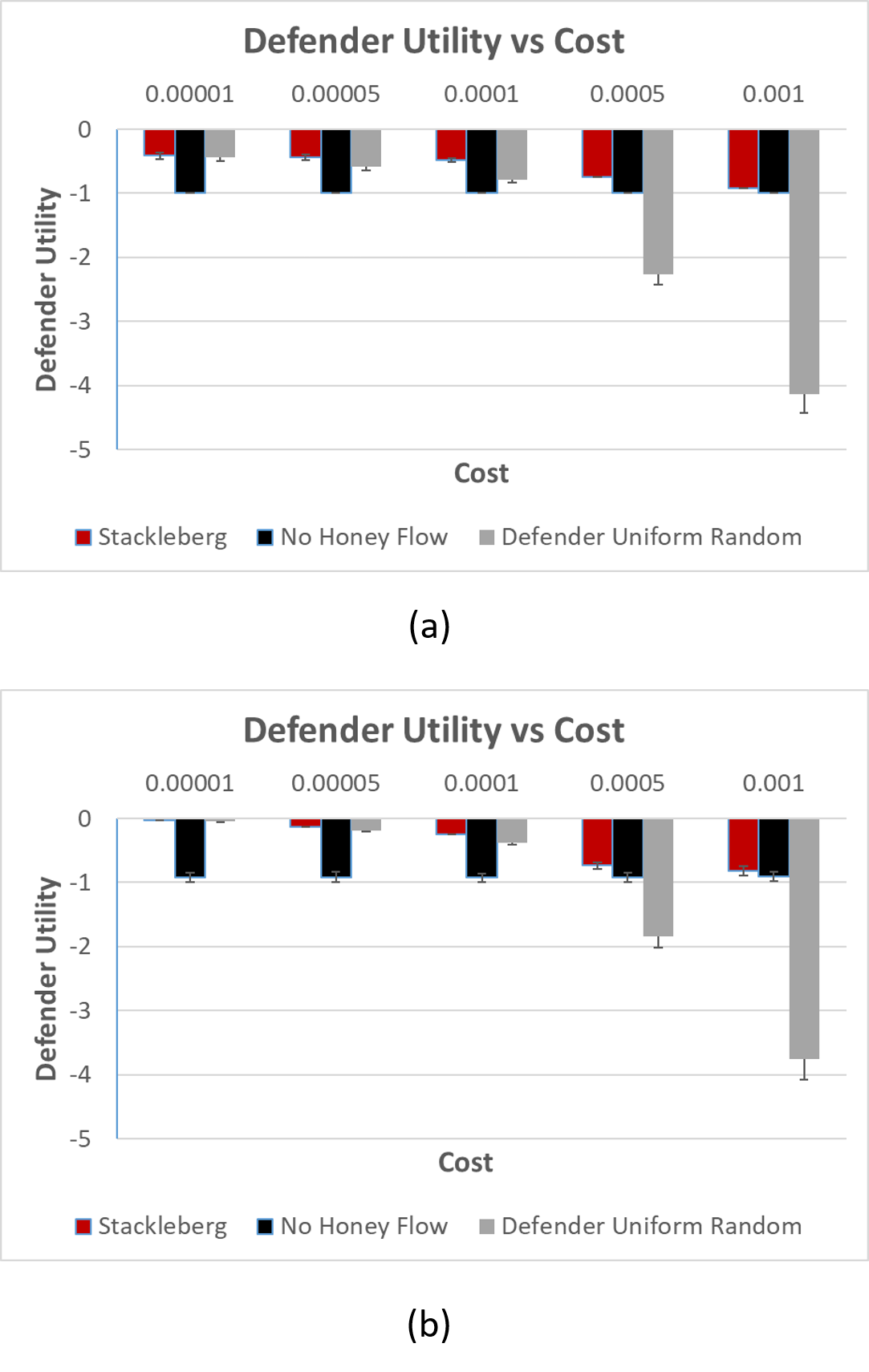}
\centering
\caption{Comparison of defender utility when the defender uses different values: a) the value of attacking a fake vulnerability is zero and a real is 1 \; b) the value of attacking a fake vulnerability is the same as real value and the values are randomly generated from [0.5, 1.0].}
\label{fig:flow-cost}
\end{figure}

Our next set of experiments focuses on evaluating the solution quality of the proposed Stackelberg game model for optimizing \emph{Snaz} honey flows compared to some plausible baselines, 1) not generating honey flows at all, 2) using a uniform random policy for generating honey flows. 
We average the results over 100 randomly generated games, each with 5 types of vulnerabilities.  
We set the number of real flows for each type to 500 and the upper bound on the number of honey flows for each type is uniform randomly generated from [500,1000]. Values are described in the caption, and we vary the costs of creating flows as shown in the Figure~\ref{fig:flow-cost}.  


The results in Figure~\ref{fig:flow-cost} show that the game theoretic solution significantly outperforms the two baselines in most settings, demonstrating the value of optimizing the honey flow generation based on the specific scenario. 
We also note that the cost has a significant impact on the overall result; with a high cost the game theories solution is similar to not generating flows at all (since they are not very cost effective). 
Random honey flow generation can be detrimental for the defender.
With a low cost, the performance of the game theoretic solution is similar to the uniform random policy; since flows are so cheap it is effective to create a very large number of them without much regard to strategy. 
With intermediate costs the value of the strategic optimization is highest, which is the most likely scenario in real applications. 

\begin{figure}[ht]
\includegraphics[width=1.0\columnwidth,height=0.5\columnwidth]{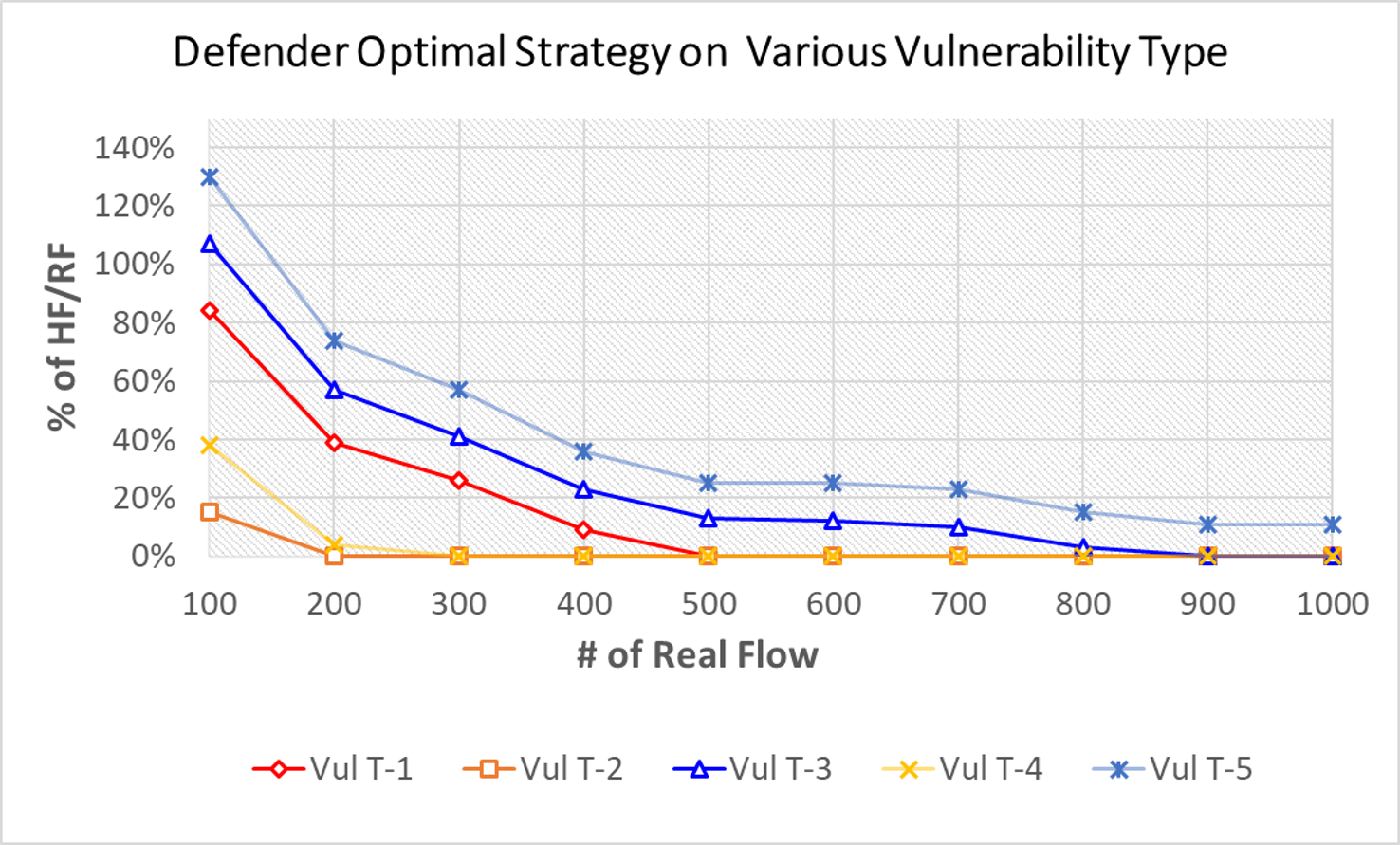}
\centering
\caption{Defender's optimal strategy as number of real flows varies.}
\label{fig:hf-rf-ratio}
\end{figure}

In our second experiment we consider vulnerabilities with different values and examine the variation in the optimal solution as we vary the number of real flows.  We use $5$ vulnerabilities with the values of the real systems ($0.8$, $0.5$, $0.9$, $0.6$, $1.0$) and attacking any fake system gives $0$. 
The cost is $0.0005$ for all types. The results in Figure~\ref{fig:hf-rf-ratio} show that the defender's strategy is to create more honey flows for the high valued vulnerabilities. As the number of real flows increases the cost of adding flows to create a high ratio is substantial and the overall number created drops for all types.  

\subsection{Solution Analysis}

\begin{figure}[htb!]
  \centering
\includegraphics[width=1\columnwidth,height=0.5\columnwidth]{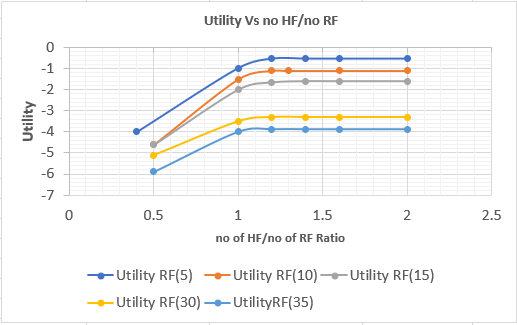}
\caption{Utility with varying honey flow ratios}
\label{fig:util}
\end{figure}

We now analyze how the ratio of honey flows to real flows changes in the optimal solution as we change the number of real flows. In Figure~\ref{fig:util}, the network setup consists of four vulnerabilities with values of (10, 20, 30, 40) and fake flows with values of (9, 18, 27, 32). The cost of generating each honey flow is 0.1.  We show the defender's expected utility as we increase the ratio of honey flows to real flows. Each line represents a different number of real flows in the original game. We see that the defender utility increases as we add honey flows, but only up to a point; when the marginal value is less than the cost the optimal solution is to stop adding additional flows.  We see this in the shape of the curves.  

\begin{figure}[htb]
\centering
\includegraphics[width=1\columnwidth,height=0.5\columnwidth]{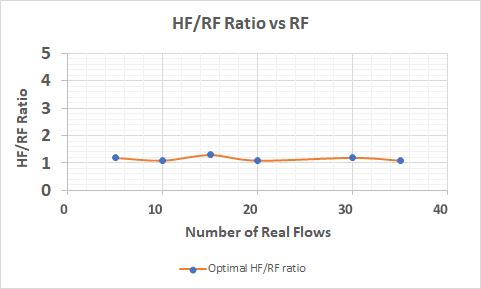}
\caption{Optimal ratios with varying real flow}
  \label{fig:util2}
\end{figure}

We note that the point where these curves flatten out is similar across all of the different numbers of real flows.  We can also see this more clearly in Figure~\ref{fig:util2}.  This suggests that the optimal strategy is relatively insensitive to the number of real flows if we consider the ratio between honey and real flows as the solution. This means that our solutions are robust to changes in the number of real flows, and that we can approximate a solution very quickly without recomputing the full solution as the number of real flows changes. 

\subsection{Scalability Evaluation}

In a practical application of \emph{Snaz} we would need to be able to calculate the optimal strategy quickly, since the network may change frequently leading to different game parameters. 
For example, the number of real flows will change over time, as will the hosts in the network.
The values of traffic and vulnerabilities, and the specific vulnerabilities we are most interested in can change also (e.g., due to the discovery of new vulnerabilities). 
We evaluate the scalability of the basic LP solution for this game as we increase the size of the game in two key dimensions: 1) increasing the number of vulnerability types, and 2) increasing the number of flows. 

\begin{figure}[htb!]
\includegraphics[width=1.0\columnwidth,height=1.1\columnwidth]{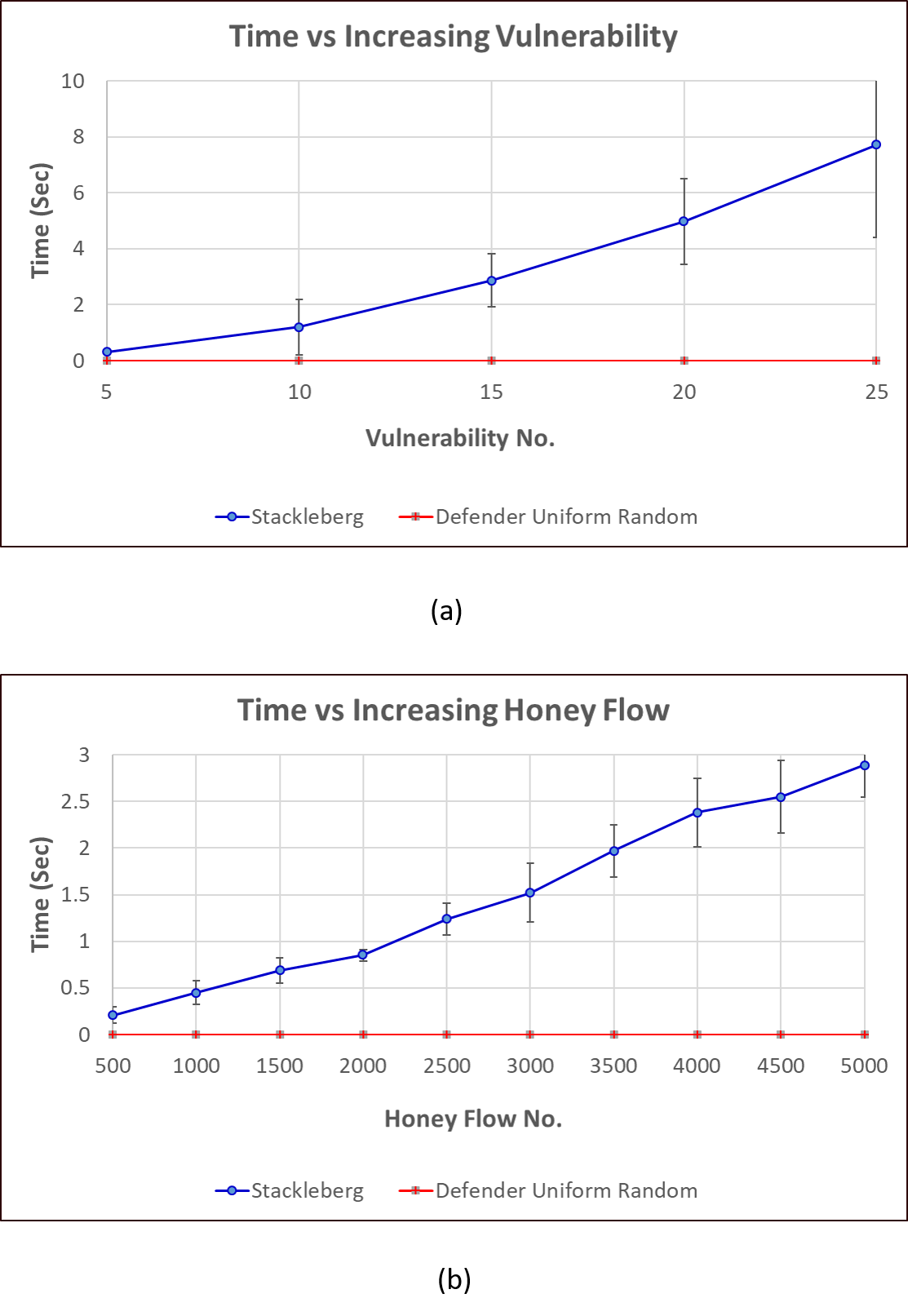}
\centering
\caption{Comparison of computational time when a) varying the number of vulnerabilities; b) varying the number of honey flows}
\label{fig:time-complexity}
\end{figure}

We randomly generate games holding the other parameters constant to evaluate the solution time. 
The results are shown in Figure~\ref{fig:time-complexity}. 
Though the solution time increases significantly as we increase the complexity of the game, we were able to solve realistic size games with a large number of flows and vulnerability types of interest within just a couple of seconds using this solution algorithm. 
This signifies that we can apply this to optimize honey flows in realistic size networks with a fast response rate; with further optimization we expect that the scalability could be improved significantly beyond this basic algorithm.  

\section{Conclusion}
\label{sec:Conclusion}

We introduced \emph{Snaz}, a technique that uses deceptively crafted honey traffic to confound the knowledge gained by adversary through passive network reconnaissance. 
We defined a Stackelberg game model for optimizing one of the key elements of \emph{Snaz}, the quantity and type of honey flows to create. This model balances cost and value trade-offs in the presence of a sophisticated attacker, but can still be solved fast enough to be used in a dynamic network environment. 
We have evaluated this model in both a preliminary emulation, as well as in simulations that explore the properties of the game theory solutions. 

There are a number of ways that this model could be improved, including incorporating network structure and variable host values into the analysis, and allowing for more overlap between vulnerabilities and flows (e.g., flows with more than one vulnerability). We can consider additional types of honey traffic, such as modifying real flows in deceptive ways. However, these may come with significantly increased computational costs for finding solutions, so we will need to develop faster algorithms to make solving these more complex models practical for a real implementation.

\section{Acknowledgement}
This work was supported by the Army Research Office under award W911NF-17-1-0370. 

\bibliographystyle{aaai}

\bibliography{AICS_2020_Mian}
\end{document}